\documentclass[aps,prd,twocolumn,groupedaddress]{revtex4}
\usepackage{graphicx}

\begin{document}

\newcommand{\kk}{\kappa}
\newcommand{\ff}{f}

\title{ Independent pair parton interactions-model of hadron interactions}

\author{I.M. Dremin}\email{dremin@lpi.ru} 
\author{V.A. Nechitailo}
\email{nechit@lpi.ru}%
\affiliation{%
Lebedev Physical Institute, Moscow
}%

\date{\today}

\begin{abstract}
A model of independent pair parton interactions is proposed,
according to which, hadron interactions are represented by a set of 
independent binary parton collisions. The final multiplicity distribution 
is described by a convolution of the negative binomial distributions in 
each of the partonic collisions. As a result, it is given by a weighted sum of 
negative binomial distributions with parameters multiplied by the number 
of active pairs. Its shape and moments are considered. Experimental data 
on multiplicity distributions in high energy $p\bar p$ processes are well 
fitted by these distributions. Predictions for the CERN Large Hadron Collider 
and higher energies are presented. The difference between $e^+e^-$ and 
$p\bar p$ processes is discussed.
\end{abstract}



\maketitle

\section{Introduction}

Hadron interactions used to be considered as proceeding via collisions of
their constituent partons. In preparton times, their role was played by pions, 
and the one-meson exchange model \cite{dche} dominated. Pions were treated as 
hadron constituents. Their high energy interaction produced a ladder of 
one-pion $t$-channel exchanges with blobs of low energy pion-pion 
interactions. This is the content 
of the multiperipheral model. These blobs were first interpreted as 
$\rho$ mesons \cite{afst} and later called fireballs \cite{acdr}, clusters 
\cite{ddun} or clans \cite{gvh} when higher mass objects were considered. 
Multiperipheral dynamics tells us that the number of these blobs is distributed 
according to the Poisson law. It was argued that its convolution with the 
distribution of the number of pions produced in each center can lead to a 
negative binomial distribution (NBD) of created particles first introduced
in \cite{giov}. This supposition 
fits experimental data on multiplicity distributions of $pp$ reactions at 
tens of GeV quite well. However, at higher energies this fit by a single NBD 
becomes unsatisfactory. A shoulder appears at high multiplicities. It is quite 
natural to ascribe it to multiple parton-parton collisions 
\cite{lpol, hump, e735, walk}, which could lead, e.g., to two-, three-, and so on, 
ladder formation \cite{cstt, kter, mwal}, and/or to different (soft, hard) types 
of interactions \cite{ghal, gugo}. They become increasingly important as the 
collision energy is increased. Better fits are achieved at 
the expense of a larger number of adjustable parameters.

This shortcoming can be minimized if one assumes that each of the high energy 
binary parton collisions is independent of others proceeding simultaneously.
With this supposition, the whole process is described as a set of independent 
pair parton interactions (the IPPI model). In fact, we assume democracy in 
sharing the initial energy of colliding hadrons among their constituents. 
The effective multiplicity of particles produced by a pair of initial partons 
does not depend on how many other
pairs interact or on what these interacting partons are (quarks or gluons). 
While parton energies vary widely at a given hadron energy, the mean
amount of energy involved in each parton-parton collision can be
comparable. Therefore, we hope 
that this simplification is valid at very high energies for such global 
characteristics as multiplicity distributions which result from an average 
over the whole phase space. If necessary, this supposition 
can be relaxed by introducing the parton distribution functions (PDF).
The inclusive distributions would call for a more detailed 
description. The IPPI model does not imply that there are no correlations
between particles. They are intrinsic in each binary collision and in their 
convolution. Surely, further correlations between these interacting pairs
of partons, of both dynamical and kinematical origin, can be introduced.
Nevertheless, the simplest model with minimum parameters and its most general 
characteristics such as multiplicity distributions should be treated first. 

\section{IPPI model and multiplicity distributions}

For multiplicity distributions, we suppose that a parton-parton
collision gives rise to a negative binomial distribution of its products 
with parameters independent of the colliding partons because the energy
is equally shared between them. Therefore, the resulting distribution of independent 
pair parton interactions is given by a sum of convolutions of processes 
with different number of participating pairs of partons weighted according 
to their probabilities. Thus one can write
\begin{eqnarray}
P(n; m, k)=\sum_{j=1}^{j_{max}}w_jP_j(n; m, k)
\nonumber \\
=\sum_{j=1}^{j_{max}}w_j\sum_{(n_p)}
\prod_{p=1}^jP_{NBD}(n_p; m, k).
\label{pn}
\end{eqnarray}
Here, $P(n; m, k)$ is the probability of creating $n$ particles, which depends on
the parameters of the NBD distribution $m$ and $k$, $n_p$ is the number 
of particles produced by the $p$th pair, $w_j$ is the probability for the $j$th pair 
to be active and $j_{max}$ is a number of active pairs. Therefore, the following
equations are valid
\begin{equation}
\sum_{p=1}^{j}n_p=n, \qquad   \sum_{j=1}^{j_{max}}w_j=1. \label{cond}
\end{equation}
The symbolical notation $\sum_{(n_p)}$ means the convolution of NBD expressions
subject to the first equation in (\ref{cond}), i.e., the sum must be taken only over 
those parton collisions whose multiplicities $n_p$ sum up to the total 
number of produced particles $n$. The NBD shape
\begin{equation}
P_{NBD}(n_p; m, k)=\frac{\Gamma(n_p+k)}{\Gamma(n_p+1)\Gamma(k)}\left(\frac{m}{k}
\right)^{n_p}\left(1+\frac{m}{k}\right)^{-n_p-k}  \label{pnbd}
\end{equation}
is characterized by two parameters $m$ and $k$, corresponding to the mean 
multiplicity 
and the dispersion $D_1$ of the distribution for a single interaction,
\begin{equation}
k^{-1}=(D_1^2-m)/m^2. \label{k}
\end{equation}
It is our supposition that, weighted by the parton distribution functions, 
such an interaction at a fixed parton-parton energy leads to a NBD. This is based
on the success of low energy fits. In multiperipheral-type models with a
Poisson distribution of created blobs, this would imply a gamma distribution
of the decay products of these blobs.

For example, the formula (\ref{pn}) explicitly 
written for three active parton pairs is as follows
\begin{widetext}
\begin{eqnarray}
P(n; m, k)=\frac{1}{\left(1+k/m\right)^n\left(1+m/k\right)^k\Gamma(k)}
\left( w_1\frac{\Gamma(n+k)}{\Gamma(n+1)}+
w_2\frac{1}{\Gamma(k)\left(1+m/k\right)^k}
\sum_{n_1=0}^n\frac{\Gamma(n_1+k)\Gamma(n-n_1+k)}{\Gamma(n_1+1)
\Gamma(n-n_1+1)}\right.
\nonumber \\
\left.
+w_3\frac{1}{\Gamma^2(k)\left(1+m/k\right)^{2k}}\sum_{n_1=0}^n
\frac{\Gamma(n_1+k)}{\Gamma(n_1+1)}\sum_{n_2=0}^{n-n_1}\frac{\Gamma(n_2+k)
\Gamma(n-n_1-n_2+k)}{\Gamma(n_2+1)\Gamma(n-n_1-n_2+1)}+\cdots\right) \label{4par}
\end{eqnarray}
\end{widetext}
Each of the three terms in this sum represents a negative binomial 
distribution because
\begin{equation}
\sum_{n_1=0}^n\frac{\Gamma(n_1+k)\Gamma(n-n_1+lk)}{\Gamma(n_1+1)\Gamma(n-n_1+1)}
=\frac{\Gamma(k)\Gamma(lk)\Gamma(n+(l+1)k)}{\Gamma((l+1)k)\Gamma(n+1)}. \label{gam}
\end{equation}
This is a general remarkable property of negative binomial distributions: 
their convolutions
result again in NBD functions with parameters multiplied by the number of
convolutions. Thus Eq. (\ref{pn}) can be rewritten as follows
\begin{equation}
P(n; m, k)=\sum_{j=1}^{j_{max}}w_jP_{NBD}(n; jm, jk).   \label{pnj}
\end{equation}
This is the main equation of the IPPI model. One gets a sum of negative
binomial distributions with shifted maxima and larger widths for a larger 
number of collisions. No new adjustable parameters appear in the distribution 
for $j$ pairs of colliding partons. All parameters are expressed in terms of
the products of parameters for a single collision and the number of collisions.
Both the mean multiplicity and dispersion $D_j^2$ for the process with $j$ active 
parton pairs are proportional to $j$.
In the total multiplicity distribution, the distributions for collisions of $j$ 
pairs of partons, $P_{NBD}$, are just weighted with their probabilities $w_j$, 
which are determined by collision dynamics and, in principle, can be 
evaluated if some model is adopted (e.g., see \cite{cstt, kter, mwal}).

An increase in the number of interacting pairs of partons in the IPPI model with 
energy gives rise to more probabilities $w_j$ different from zero. Certainly,
all the parameters $w_j, m, k$ depend on energy. This dependence is implied 
but not shown explicitly in the above formulas. 
One can hope that at 
asymptotically high energies the probability for $j$ pairs of independent 
interactions $w_j$ is the product of $j$ probabilities $w_1$ for one pair,
\begin{equation}
w_j=w_1^j. \label{wjw1}
\end{equation}
From the normalization condition 
\begin{equation}
\sum_{j=1}^{j_{max}}w_j=\sum_{j=1}^{j_{max}}w^j_1=1,   \label{w1w1}
\end{equation}
one can find $w_1$ if $j_{max}$, which is determined by the maximum number of
parton interactions at a given energy, is known. In fact, the value of $w_1$ 
ranges between 1 at low energies (for $j_{max}$=1) and 0.5 at asymptotics where
$j_{max}$ tends to infinity. With energy increase, it approaches the second value 
from above, passing through some thresholds, and is already quite close to it 
at the present highest energies. 

Thus we are left with only two parameters of the model,
$m$ and $k$, which can be found from fits of experimental data. 
The dependence on the number of collisions $j_{max}$ and on the probabilities 
$w_j$ is determined by the behavior of the moments of probabilities
\begin{equation}
M_r=\sum_{j=1}^{j_{max}}w_j j^r,  \label{mr}
\end{equation}
as explicitly shown in the Appendix for ranks $r\leq 5$. In particular, 
the average multiplicity is given by
\begin{equation}
\langle n\rangle=mM_1.    \label{nav}
\end{equation}
If one assumes some extrapolation of $\langle n\rangle$ to higher energies, 
it can be used for prediction of the distributions. Let us emphasize that the 
asymptotic behavior of $m$ is directly related to that of the mean multiplicity,
\begin{equation}
\langle n_{as}\rangle=m_{as}\sum_{j=1}^{\infty}jw_1^j=m_{as}\frac{w_1}
{(1-w_1)^2}=2m_{as}.    \label{mas}
\end{equation}
The value of $m$ is usually quite close to the position of the maximum of the
distribution. Thus the relation (\ref{mas}) tells us that in the IPPI model the
asymptotic mean multiplicity is about twice larger than the location of its
maximum determined mainly by a single parton-parton interaction.
One can expect that the
asymptotic relation for the probabilities (\ref{wjw1}) becomes valid only at
energies where four or more pairs are already active.
This series becomes a polynomial at finite energies. In practice, 
the threshold effects should also be taken into account at finite energy. They 
would somewhat suppress $w_j$ at the largest $j$ and, correspondingly, enlarge 
the role of one- and two-pair interactions.

In \cite{mwal}, the energy dependence of the probabilities $w_j$ 
was estimated according to the multiladder exchange model \cite{kter} but
Poisson distributions were used for each of the ladders. 
The probabilities are given by the following normalized expressions
\begin{equation}
w_j(\xi_j)=\frac{p_j}{\sum_{j=1}^{j_{max}}p_j}
=\frac {1}{jZ_j\left(\sum_{j=1}^{j_{max}}p_j\right)}
\left( 1-e^{-Z_j}\sum_{i=0}^{j-1}\frac {Z_j^i}{i!}\right)      \label{kter}
\end{equation}
where 
\begin{equation}
\xi_j=\ln(s/s_0j^2), \;\;  Z_j=\frac {2C\gamma}{R^2+\alpha_P'\xi_j}
\left(\frac{s}{s_0j^2}\right)^{\Delta}  \label{xiz}
\end{equation}
with numerical parameters obtained from fits of the
experimental data on total and elastic scattering cross sections: 
$\gamma=3.64$ GeV$^{-2}$, $\,$ $R^2=3.56$ GeV$^{-2}$,
$\,$ $C=1.5,$ $\,$ $\Delta=\alpha_P-1=0.08,$ $\,$ $\alpha_P'=0.25$ GeV$^{-2}$,
$s_0$ =1 GeV$^2$. 

Below, we will use both possibilities (\ref{wjw1}) and 
(\ref{kter}) in our attempts to describe the experimental data.
The probabilities $w_j$ are different for each (see Table~\ref{tab:table1}).
In the IPPI model they decrease exponentially with increasing number of active partons,
while in the ladder model they are inversely proportional to this number
with additional suppression at large $j$ due to the term in parentheses in
(\ref{kter}). This is the result of the modified eikonal approximation.

We show the values $w_j$ for 3--6 pairs calculated 
according to Eq. (\ref{wjw1}) in the left-hand side of Table~\ref{tab:table1} 
and according
to Eq. (\ref{kter}) in its right-hand side. These values 
$j_{max}$ are chosen because they will be used in the comparison with experiment.
In particular, we shall choose $j_{max}=3$ at 300 and 546 GeV, 4 at 1000 and
1800 GeV, 5 at 14 TeV, and 6 at 100 TeV (see below).

\begin{table}

\caption{\label{tab:table1}
The values of $w_j$ according to Eq. (\ref{wjw1}) (left-hand side) and
Eq. (\ref{kter}) (right-hand side).}

\begin{tabular}{|c||c|c|c|c||c|c|c|c|} \hline
$j_{max}$& 3&  4&     5&     6   &    3 &    4 &    5 &     6  \\ \hline
$w_1$& 0.544& 0.519& 0.509& 0.504& 0.562& 0.501& 0.450& 0.410\\ 
$w_2$& 0.295& 0.269& 0.259& 0.254& 0.278& 0.255& 0.236& 0.219\\
$w_3$& 0.161& 0.140& 0.131& 0.128& 0.160& 0.153& 0.152& 0.147\\
$w_4$& 0    & 0.072& 0.067& 0.065& 0    & 0.091& 0.100& 0.104\\ 
$w_5$& 0    & 0    & 0.034& 0.033& 0    & 0    & 0.062& 0.073\\
$w_6$& 0    & 0    & 0    & 0.016& 0    & 0    & 0    & 0.047\\ \hline
\end{tabular}                    
\end{table}

One can clearly see the difference between the two approaches. The value of $w_1$
is always larger than 0.5 in the IPPI model while it can become less than 0.5 
in the ladder model \cite{cstt,kter} at high energies. 
In the ladder model, $w_j$ depend explicitly on energy (not only on the $j_{max}$ 
cutoff). We show their 
values at 546 and 1800 GeV in the right-hand side columns of $j_{max}$=\,3 
and 4. Those at 300 and 1000 GeV are larger for $w_1$ by about 1$\%$ and smaller 
for $w_3$ by about 3$\%$.  When the energy increases, the processes with 
a larger number of active pairs play a more important role in the ladder approach 
compared to the IPPI model. Thus, the $j_{max}$ cutoff is also more essential there.

In principle, one can immediately try a two-parameter fit of experimental
multiplicity distributions using Eq. (\ref{pnj}) if $w_j$ are known.
However, the use of their moments is preferred as shown below.

\section{Moments of multiplicity distributions}

The shapes of the multiplicity distributions $P(n)$ usually look quite complicated.
Often, they are better represented by their moments, which also contain complete
information. The easiest way to define them is to introduce the
generating function
\begin{equation}
G(z) = \sum_{n=0}^{\infty }P(n)(1+z)^{n}.                     \label{3}
\end{equation}
In what follows, we will use the so-called unnormalized factorial ${\cal F}_q$
and cumulant ${\cal K}_q$ moments defined according to the formulas
\begin{equation}
{\cal F}_{q} = \sum_{n} P(n)n(n-1)...(n-q+1) =
 \frac {d^{q}G(z)}{dz^{q}}\vline _{z=0}, 
\label{4}
\end{equation}
\begin{equation}
{\cal K}_{q} = \frac {d^{q}\ln G(z)}{dz^{q}}\vline _{z=0}. \label{5}
\end{equation}
They correspondingly determine the total and genuine correlations among 
the particles produced (for more details, see \cite{ddki}). For $q=1$, they 
define the mean multiplicity; the second moment is related to the 
width of the distribution, etc. The factorial moments are evaluated from 
experimental data according to their definition (\ref{4}).
Both ${\cal F}_{q}$ and 
${\cal K}_{q}$ grow, however, extremely fast with their ranks. Therefore, 
it is more convenient to use \cite{13} their ratio 
$H_q={\cal K}_{q}/{\cal F}_{q}$, where these dependencies partly cancel. 
This ratio is easy to find from iterative formulas:
\begin{equation}
H_q=1-\sum_{p=1}^{q-1}\frac{\Gamma(q)}{\Gamma(p+1)\Gamma(q-p)}H_{q-p}\frac
{{\cal F}_p{\cal F}_{q-p}}{{\cal F}_q},   \label{hqfq}
\end{equation}
once the factorial moments have been evaluated. Thus, both ${\cal F}_q$ and
$H_q$ are determined by experimental data according to Eqs. (\ref{4}) and 
(\ref{hqfq}).

Recall that these ratios appear quite naturally in QCD as the solutions 
of the equations for the generating functions of multiplicity distributions 
\cite{13}. Therefore, their use is especially 
informative because one can compare experimental results and model
calculations with analytical QCD predictions for jets in $e^+e^-$ 
annihilation as reviewed in \cite{dgar}. QCD predicts a very specific behavior 
of the $H_q$ moments as functions of ranks $q$ and energy. It has been shown 
\cite{13,41} that $H_q$ for jets at the present energies (SLC, LEP) should 
oscillate, and this prediction has been confirmed by experimental results 
\cite{dabg, sld, l3}. The first minimum is located near $q=5$ at Z$^0$ energy.
At higher energies, this minimum moves to larger values of $q$; the oscillations 
become less pronounced and disappear in the asymptotics where $H_q=1/q^2$. 
Moreover, these oscillations have been found \cite{dabg,dgar} even for 
experimentally studied $pp$ and $AA$ collisions. In these cases, they 
can be ascribed to the multicomponent structure of the processes. Such 
a structure is incorporated in the IPPI model according to Eq. (\ref{pn}).
We will see if it is enough to describe experimental data. 

Let us emphasize that
$H_q$ moments are very sensitive to minute details of the multiplicity distributions
and can be used to distinguish between different models and experimental data.
However, one should be warned that the amplitudes of the oscillations strongly
depend on the multiplicity distribution cutoff due to limited experimental
statistics (or by another reasoning) if the experiment is done at rather low 
multiplicities. There are no cutoffs in analytical expressions for $H_q$. 
One can control the influence of cutoffs by shifting them appropriately. 
The qualitative features persist nevertheless. In what follows, we consider
very high energy processes where the cutoff due to experimental statistics
is practically insignificant.

The IPPI model predicts new special features of the moments ${\cal F}_{q}$ and 
$H_q$. The factorial moments of the distribution (\ref{pnj}) are
\begin{equation}
{\cal F}_{q} = \sum_{j=1}^{j_{max}}w_j\frac{\Gamma(jk+q)}{\Gamma(jk)}
\left(\frac{m}{k}\right)^q=f_q(k)\left(\frac{m}{k}\right)^q   \label{fphi}
\end{equation}
with
\begin{equation}
f_q(k)=\sum_{j=1}^{j_{max}}w_j\frac{\Gamma(jk+q)}{\Gamma(jk)}=
k\sum_{j=1}^{j_{max}}w_jj(jk+1)\cdots(jk+q-1). \label{phiq}
\end{equation}
The cumulant moments are written as
\begin{equation}
{\cal K}_{q} =\kappa_q(k)\left(\frac{m}{k}\right)^q.  \label{ckap}
\end{equation}
The explicit $k$ dependence of $f_q(k)$ and $\kappa_q(k)$ for $q\leq 5$ is
shown in the Appendix.
For $H_q$ moments one gets
\begin{equation}
H_q=1-\sum_{p=1}^{q-1}\frac{\Gamma(q)}{\Gamma(p+1)\Gamma(q-p)}H_{q-p}\frac
{f_pf_{q-p}}{f_q}.   \label{hqph}
\end{equation}
Note that according to Eq. (\ref{hqph}) $H_q$ are functions of the 
parameter $k$ only and do not depend on $m$ in the IPPI model, because 
the $m$ dependence of factorial and cumulant moments is the same. 
This remarkable property of the $H_q$ moments provides an opportunity to
fit the same results with a smaller number of parameters. If the $w_j$'s 
are given by Eq. (\ref{w1w1}), the only adjustable 
parameter left is $k$. These moments decrease with increase of $k$ and $q$. 

Once the parameter $k$ is found from fits of $H_q$, it is possible to get
another parameter $m$ by rewriting Eq. (\ref{fphi}) as follows:
\begin{equation}
m=k\left(\frac {{\cal F}_{q}}{f_q(k)}\right)^{1/q}. \label{mkfp}
\end{equation}
This formula is a sensitive test for the whole approach because it states
that the definite ratio of $q$-dependent functions to the power $1/q$ becomes
$q$ independent if the model is correct. Moreover, this statement should be 
valid only for those values of $k$ that are determined from $H_q$ fits.
Therefore, it can be considered as a criterion for a proper choice of $k$ 
and for the model validity, in general.

One substitutes the experimentally determined values of factorial moments, 
divides them by the theoretical functions $f_q$ given by Eq. (\ref{phiq}), 
and examines whether this ratio to the power $1/q$ is independent of $q$
at $k$ values found previously from $H_q$ fits. If the answer is positive, 
the parameter $m$ is known according to Eq. (\ref{mkfp}). If not, the 
model should be modified. With parameters $k$ and $m$ found, one can 
try to fit the shapes of experimentally measured multiplicity 
distributions directly. This is another test of the self-consistency 
of the IPPI model.

At the same time, the value of $m$ determines the position of the peak of 
the multiplicity distribution. For a given $\langle n\rangle$, it can be used
to check the choice of probabilities $w_j$ according to Eq. (\ref{nav}).

Recall that both parameters $m$ and $k$ depend on the energy of 
the colliding hadrons $s$. This
dependence can be determined from fits of experimentally found values of $H_q$ 
and ${\cal F}_{q}$ as explained above. To extrapolate it to higher energies, 
one should use some guesses. Since $m$ has the meaning of the average 
multiplicity of a binary parton collision, it should behave similarly to 
the mean multiplicity of the whole process. The latter is usually fitted
by a logarithmic dependence with some log-squared terms added. 
No experience has been gained yet for the parameter $k$.

The Poisson distribution possesses the same property of convolutions which made it 
possible to get Eq. (\ref{pnj}) for NBD distributions. Therefore, all the above
relations are valid for a model with convoluted Poisson distributions.
Actually, they can be obtained in the limit $k\rightarrow \infty $. 
For example, the factorial moments are ${\cal F}_{q}^{(Poisson)}=m^qM_q$.

\section{Comparison with experiment}

We have compared IPPI model conclusions with experimental multiplicity
distributions of the E735 Collaboration \cite{e73} for $p\bar p$ collisions at 
energies 300, 546, 1000, and 1800 GeV extrapolated \cite{walk,arn} to the full
phase space.
The multiplicity of charged particles was divided by 2 to get the multiplicity
of particles with the same charge. Then the above formulas for the moments were
used. Correspondingly, the parameters $m$ and $k$ refer to these distributions.

An analysis of experimental data done in \cite{walk} has shown that two parton 
pairs are already active at energies above 120 GeV. The thresholds for triple
or more parton-parton collisions are less definite. They depend on the form
of the multiplicity distribution adopted for a single collision. We assume
that three parton pairs are active at 300 and 546 GeV and four 
at 1000 and 1800 GeV
with NBDs for a single collision. We use these values in our calculations.

\begin{figure}
\resizebox{\columnwidth}{!}{\includegraphics{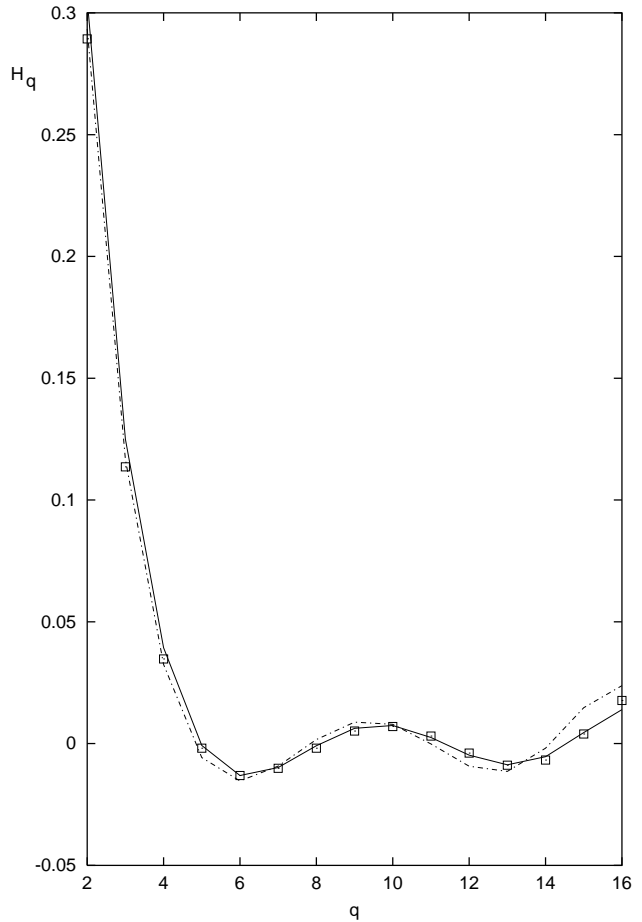}}
\caption{\label{fig:fig1}  A comparison of $H_q$ moments derived from experimental data at 1.8 TeV (squares)
         with their values calculated with parameter $k=4.4$ (dash-dotted line) and 3.7 (solid line).}
\end{figure}

Factorial and $H_q$ moments were obtained from experimental data on 
$P(n)$ according to Eqs. (\ref{4}) and (\ref{hqfq}).
Experimental $H_q$ moments were fitted by Eq. (\ref{hqph}) 
to get the parameters $k(E)$ of the IPPI model. We show in Fig. 1 how perfect are 
these fits at 1.8 TeV for $k$ equal to 3.7 (solid line) and 4.4 (dash-dotted
line). At this energy, we consider
four active parton pairs with $w_j$ given by Eq. (\ref{w1w1}) (the second column
in Table~\ref{tab:table1}). It is
surprising that the oscillations of $H_q$ moments are so well reproduced with
one adjustable parameter $k$. The general tendency of this quite 
complicated oscillatory dependence is clearly seen.

\begin{figure}
\resizebox{\columnwidth}{!}{\includegraphics{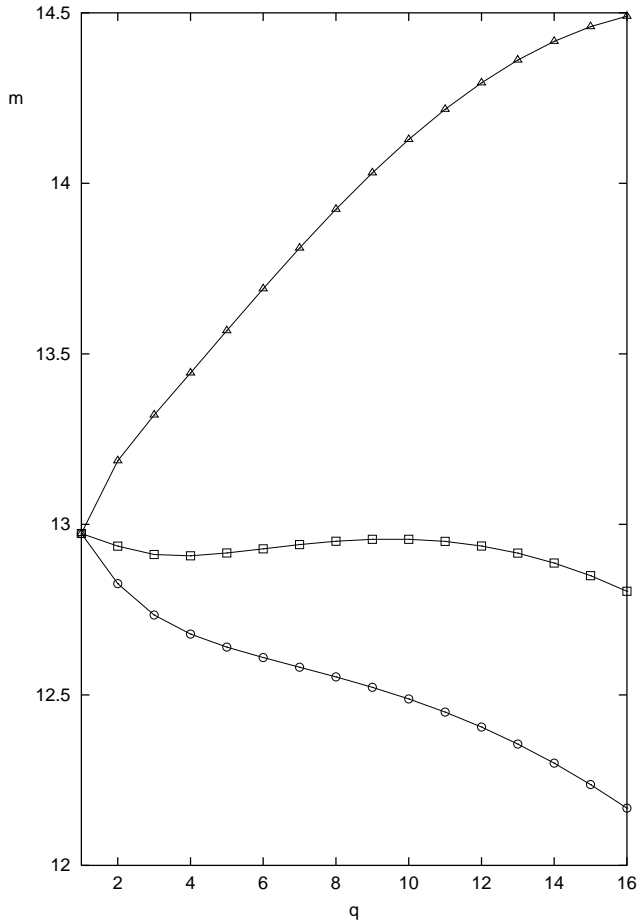}}
\caption{\label{fig:fig2}  The $q$ dependence of $m$ for $k=4.4$ (squares), 
3.7 (circles), and 7.5 (triangles).}
\end{figure}

With these values of the parameter $k$, we have checked whether $m$ is constant 
as a function of $q$. Experimental factorial moments and IPPI values for $f_q$ 
were inserted in Eq. (\ref{mkfp}). The $m(q)$ dependence is shown in Fig. 2 
for the same values of $k$=\,4.4 (squares) and 3.7 (circles) and for a 
much larger value 7.5 (triangles). The constancy of $m$ is satisfied with an accuracy
better than 1.5$\%$ for $k=4.4$ up to $q=16$. The upper and lower lines 
in Fig. 2 demonstrate clearly that this condition substantially bounds 
the admissible variations of $k$. 

\begin{figure}
\resizebox{\columnwidth}{!}{\includegraphics{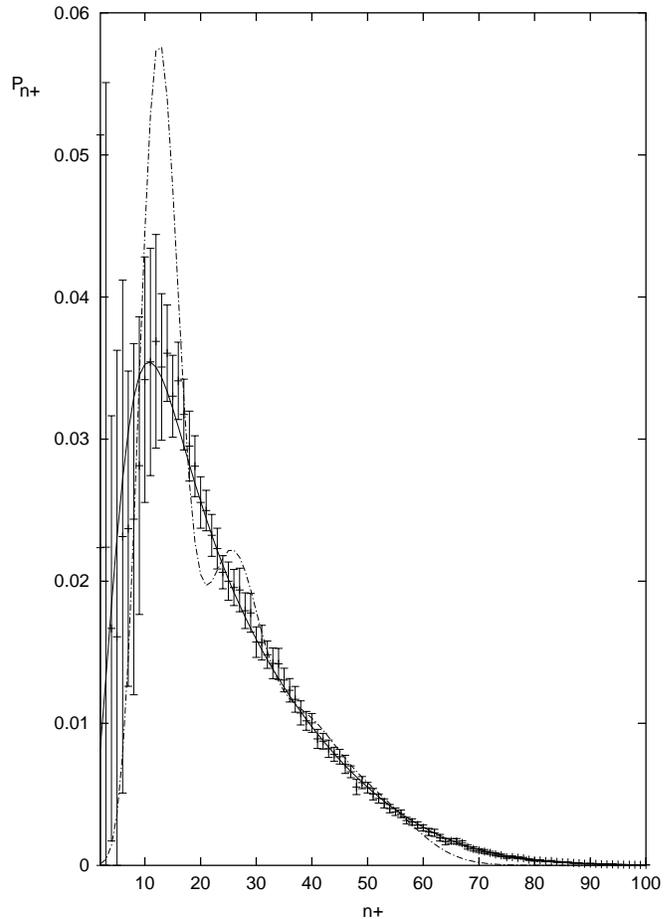}}
\caption{\label{fig:fig3}  The multiplicity distribution at 1.8 TeV and its 
fit at $m=12.94, k=4.4$ (solid line). 
 The dash-dotted line demonstrates what would happen if the NBD were replaced 
 by Poisson distribution.}
\end{figure}

It is well known that experimental cutoffs of multiplicity distributions 
due to the limited statistics of an experiment can influence the behavior of 
$H_q$ moments. Consequently, they impose some limits on the $q$ values that can 
be considered when a comparison is done. Higher rank moments can be evaluated
if larger multiplicities have been measured. To estimate the admissible range 
of $q$, we use the results obtained in QCD. Characteristic multiplicities that
determine the moment of the rank $q$ can be found. By inverting this relation,
one can write the asymptotic expression for the characteristic range 
of $q$ \cite{yuri}. This provides the bound $q_{max}\approx
Cn_{max}/\langle n\rangle $ where $C\approx 2.5527$. However, it 
underestimates the factorial moments. Moreover, the first moment is not 
properly normalized (it becomes equal to $2/C$ instead of 1). The 
strongly overestimated values (however, with a correct normalization
of the first moment)
are obtained if $C$ is replaced by 2. Hence, one can say
that the limiting values of $q$ are given by the inequalities
\begin{equation}
2n_{max}/\langle n\rangle < q_{max}\leq Cn_{max}/\langle n\rangle.  \label{qmax}
\end{equation}
The ratio $n_{max}/\langle n\rangle $ measured by the E735 Collaboration at 1.8 TeV
is about 5. Thus, $q_{max}$ should be in the interval between 10 and 13. The 
approximate constancy of $m$ and
proper fits of $H_q$ demonstrated above persist to even higher ranks.

The same-charge multiplicity distribution at 1.8 TeV has been fitted with the
parameters $m=12.94$ and $k=4.4$ as shown in Fig. 3 (solid line). To estimate
the accuracy of the fit, we calculated 
$\sum_{n=1}^{125}[P_t(n)-P_e(n)]^2/\Delta ^2$ over all 125 experimental points. 
Here, $P_t,P_e$ are the theoretical and experimental distributions and $\Delta$ is
the total experimental error. It includes both statistical and systematical
errors. Note that the latter are large at low multiplicities in the E735 data.
This sum is equal to 50 for 125 degrees of freedom. No minimization of it was 
attempted. This is twice better than the three-parameter fit by a 
generalized NBD considered in \cite{heg}.
A Poisson distribution of particles in binary collisions is completely 
excluded. This is shown in Fig. 3 by the dash-dotted line.

\begin{figure}
\resizebox{\columnwidth}{!}{\includegraphics{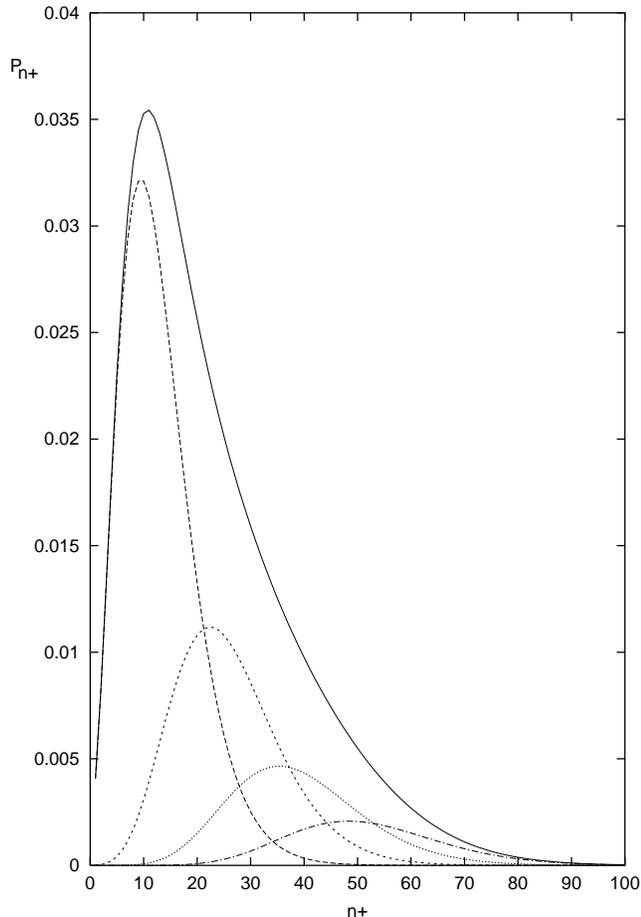}}
\caption{\label{fig:fig4}  The decomposition of the fit in 
Fig. 3 to one, two, three, and four parton-parton collisions.}
\end{figure}

We show in Fig. 4 the decomposition of the fit in Fig. 3 to processes
with different numbers $j$ of parton pairs involved in collision.
It is seen that the locations of their maxima are approximately proportional to $j$.

\begin{figure}
\resizebox{\columnwidth}{!}{\includegraphics{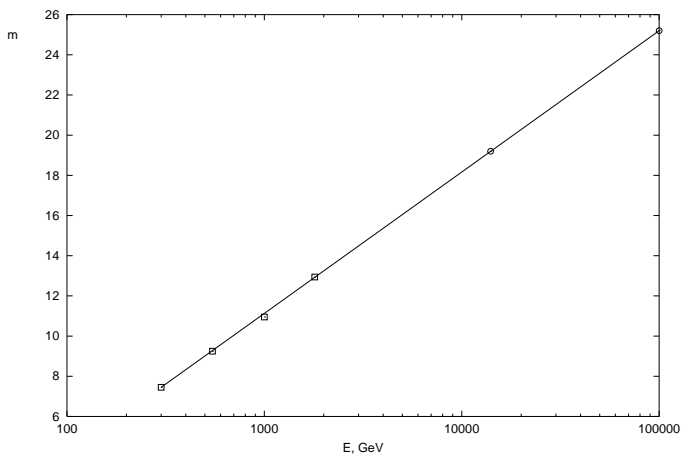}}
\caption{\label{fig:fig5} The energy dependence of $m$ (squares) and its linear 
extrapolation (circles at 14 and 100 TeV).}
\end{figure}

\begin{figure}
\caption{\label{fig:fig6} The values of $k$ as calculated with $w_j$ satisfying 
the relation (\ref{wjw1}) (squares).}
\end{figure}

The same procedure has been applied to data at energies 300, 546, and 1000 GeV. 
As stated above, we have assumed that three binary parton collisions are 
active at 300 and 546 GeV and four at 1000 GeV. We 
plot in Figs. 5 and 6 the energy dependence of the parameters $m$ and $k$. 
The parameter $m$ increases logarithmically with energy. This is expected
[see (\ref{nav})] because increase of $M_1$ due to increasing numbers of 
active pairs at these energies leads to a somewhat faster 
than logarithmic increase of the average multiplicity in accordance with
experimental observations. The energy
dependence of $k$ (crosses) is more complicated and rather irregular. 

We tried to ascribe the latter to the fact that the effective values of $k$, 
which we actually find from these fits, depend on the effective number of 
parton interactions, i.e., on the $w_j$ variation at a threshold.
The threshold effects can be important in this energy region. Then the simple 
relation (\ref{wjw1}) is invalid. This influences the functions $f_q(k)$
[Eq. (\ref{phiq})] and, consequently, $H_q$ calculated from Eq. (\ref{hqph}).
One can reduce the effective number of active pairs to about 2.5 at 300 GeV 
and 3.5 at 1000 GeV if one chooses the following values of $w_j$: 0.59,$\,$ 0.34,
and 0.07 at 300 GeV and 0.54,$\,$ 0.29,$\,$ 0.14, and 0.03 at 1000 TeV instead
of those calculated according to Eq. (\ref{wjw1}) and shown in Table~\ref{tab:table1}.
This 
gives rise to values of $k$ which are not drastically different from the previous
ones. However, the quality of the fits becomes worse.
Fits with two active pairs at 300 GeV and three pairs at 1000 GeV fail completely. 

Hence, we have to conclude that this effect results from some dynamics of the hadron
interactions that is not understood yet and should be incorporated in
the model. The preliminary explanation of this effect could be that at
the threshold of new pair formation the previous active pairs produce
more squeezed multiplicity distributions due to the smaller phase-space room 
available for them because of the newcomer. Therefore, the single pair 
dispersion decreases and the $k$ values increase. This would imply 
that thresholds are marked not only by the change of $w_j$ shown in 
Table~\ref{tab:table1}
but also by the variation of the parameter $k$.

The threshold effects become less important at higher energies. We assume 
that there are five active pairs at 14 TeV and six at 100 TeV. Then we extrapolate 
to these energies. The parameter $m$ becomes equal to 19.2 at 14 TeV and 25.2 
at 100 TeV if logarithmic dependence is adopted as shown in Fig. 5 by the
straight line. The predicted multiplicity distributions are plotted in Fig. 7.
We choose two values of $k$ equal to 4.4 (solid line) and 8 (dash-dotted line)
for 14 TeV. Low multiplicities
are suppressed at larger $k$, and the maximum is slightly shifted to higher
multiplicities. The shape of the tail is practically unchanged. For 100 TeV,
we show only the prediction for $k=4.4$ (dashed line) because increase of $k$ 
leads to the same qualitative effect as for 14 TeV.
The oscillations of $H_q$ still persist at these energies (see Fig. 8).
The minima are, however, shifted to $q=6$ at 14 TeV and 7 at 100 TeV as expected.

\begin{figure}
\resizebox{\columnwidth}{!}{\includegraphics{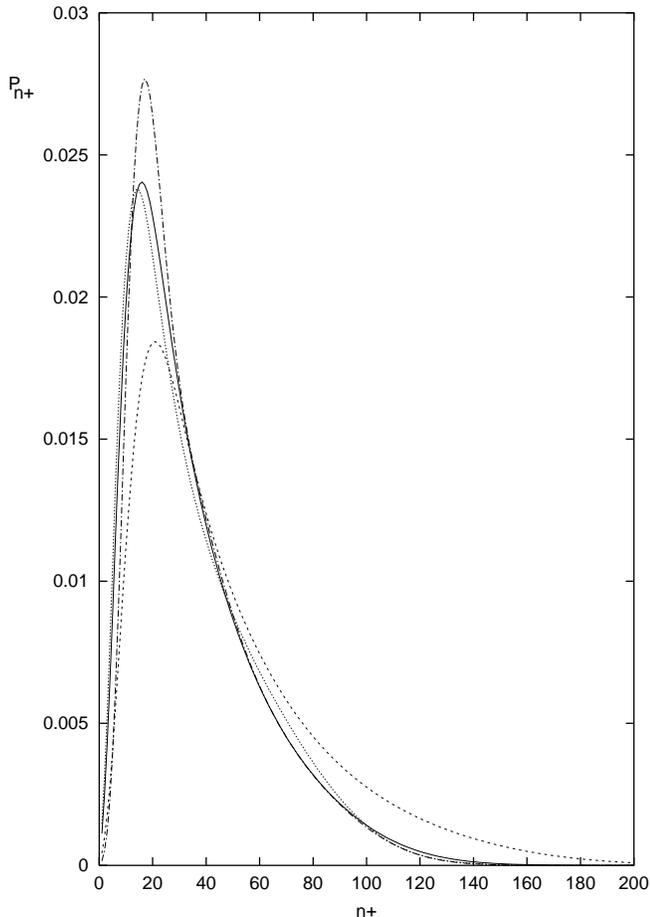}}
\caption{\label{fig:fig7} The same-charge multiplicity distributions at 14 TeV and 100 TeV obtained by 
         extrapolation of parameters $m$ and $k$ with five active pairs at 14 TeV and six at 100 TeV
         (for the IPPI model: solid line, 14 TeV, $k=4.4$; dash-dotted line 14 TeV, $k=8$; 
         dashed line 100 TeV, $k=4.4$; for the ladder model: dotted line 14 TeV, $k=4.4$).
}
\end{figure}

\begin{figure}
\resizebox{\columnwidth}{!}{\includegraphics{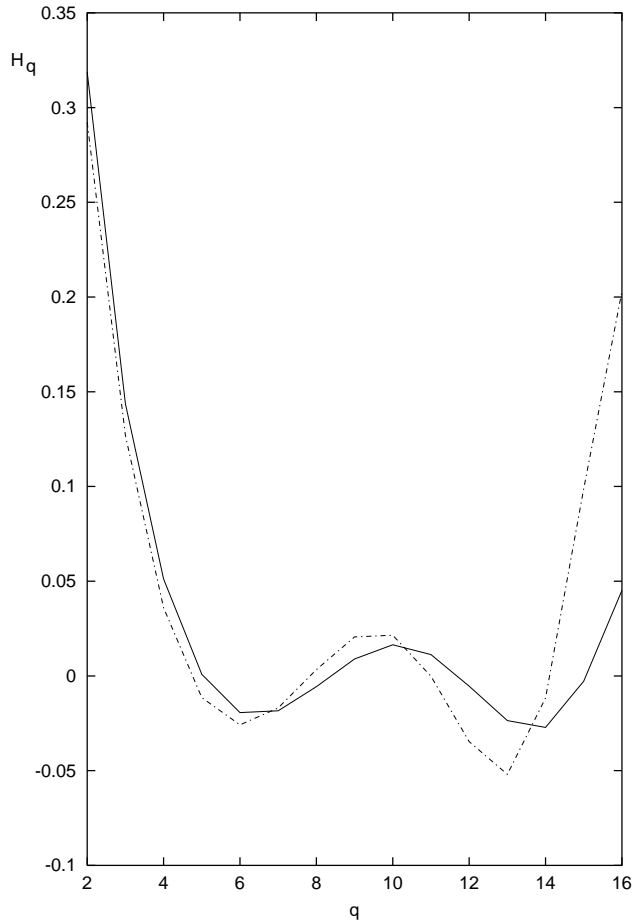}}
\caption{\label{fig:fig8} The behavior of $H_q$ predicted at 14 TeV 
($k=4.4$, solid line; $k=8$, dash-dotted line).}
\end{figure}

The fit at 1.8 TeV with an approximation of $w_j$ according to the ladder model
(\ref{kter}) with a NBD for a binary parton collision is almost as successful as 
the fit with values of $w_j$ given by the IPPI model. However, some difference 
at 14 TeV between these models is predicted (compare the solid and dotted lines in 
Fig. 7, both obtained for $k=4.4$). This difference becomes more pronounced at 
100 TeV. To keep the same mean multiplicity in both models at the same energy, 
we have chosen different values of $m$ as dictated by Eq. (\ref{nav}) and
the $w_j$ values shown in Table~\ref{tab:table1}; 
namely, their ratios are $m_{IPPI}/m_{lad}$
=\,0.988, 1.039,  1.123, 1.228 for $j_{max}$=3,  4,  5, 6, respectively. 
This shows that the maximum of the distribution 
moves to smaller multiplicities and its width becomes larger in the ladder
model compared to the IPPI model with energy increase.

Certainly, one should not overestimate the success of the IPPI model in its present
initial state. It has been applied just to multiplicity distributions. 
For more detailed properties, say, rapidity distributions, one would
need a model for the corresponding features of the one-pair process.
Moreover, the screening effect (often described by the triple Pomeron vertex)
will probably become more important at higher energies. All these features
are implemented in some way in the well known Monte Carlo programs PYTHIA 
\cite{pyth}, HERWIG \cite{herw} and DPM-QGSM \cite{cstt,kter}. However, for 
the last one, the multiplicity
distribution for a single ladder is given by the Poisson distribution of 
emission centers (resonances) convoluted with their decay properties,
and the probabilities $w_j$ contain several adjustable parameters. It differs
from the IPPI model. The present 
approach proposes more economic way with a smaller number of such parameters.
Concerning the further development of event generator codes, it is 
tempting to incorporate there the above approach with a negative binomial
distribution of particles created by a single parton pair, and confront the 
results with a wider set of experimental data. This has not been done
yet for the IPPI model, and we intend to work on it
later to learn how it influences other characteristics.

It would also be interesting to see whether this model is valid for AA collisions 
as well whether the collective effects (saturation?) prevent its application there.
This work is in progress now.

\section{Are $e^+e^-$ and $p\bar p$ similar?}

This question was raised by a recent statement of the PHOBOS Collaboration \cite{phob}
that the energy behavior of mean multiplicities in all processes is similar.
It was concluded that the dynamics of all hadronic processes is the same.
In addition to our general belief in QCD, we cannot claim that other characteristics
of multiple production processes initiated by different partners coincide.

At first sight, QCD fits of multiplicity distributions in $e^+e^-$ collisions
and IPPI model fits of $p\bar p$ collisions are completely unrelated and cannot 
be compared. There is, however, one definite QCD prediction that
allows us to ask the question whether QCD and the IPPI model are compatible. 
This is the asymptotic behavior of $H_q$ moments in QCD. They should behave
\cite{13} as $H_q^{as}=1/q^2$. One can also determine the asymptotics of
$H_q$ moments in the IPPI model and compare both approaches. The
asymptotical values of the probabilities $w_j$ (\ref{wjw1}) and their moments
$M_r$ (\ref{mr}) for $r\leq 5$ are as follows:
\begin{eqnarray}
w_j=0.5^j, \; M_1=2, \; M_2=6, \; M_3=26, \nonumber\\ 
M_4=150, \; M_5=1082. \label{wm}
\end{eqnarray}
Inserting them in the expressions for $\kappa_q$ and $f_q$ given in the Appendix,
one can evaluate the asymptotic behavior of the $H_q$ moments in the IPPI model 
at any parameter $k$. All asymptotic $H_q$ are decreasing functions of $k$. Their 
minimum values are reached at $k\rightarrow \infty$, i.e., for a convolution of
Poisson distributions. They are given by the ratio of the coefficients in 
front of the leading $k^q$ terms in $\kappa_q$ and $f_q$ (see the Appendix) and are 
equal to
\begin{equation}
H_2^{(P)}=\frac{1}{3}, \; H_3^{(P)}=\frac{3}{13}, \; 
H_4^{(P)}=\frac{13}{75}, \; H_5^{(P)}=\frac{75}{541}. \label{has}
\end{equation}
These values are noticeably larger than QCD predictions of $1/q^2$. Since they
are even larger for any finite parameter $k$, we have to state that QCD and
the IPPI model have different asymptotics. In other words, this implies that
Eq. (\ref{hqph}), considered as an equation for $f_q$ with $H_q=1/q^2$ 
inserted in it, does not have a solution with asymptotical values of $M_r$ 
[Eq. (\ref{wm})] in the IPPI model.

It is an open question whether other
asymptotic relations for $w_j$ different from Eq. (\ref{wjw1}) can be found which
would lead to the same behavior of $H_q$ moments in $p\bar p$ and $e^+e^-$ 
collisions, i.e., if a solution of Eq. (\ref{hqph}) can be found for some
values of $M_r$ different from those given by Eq. (\ref{wm}). Only then one can 
hope to declare an analogy between these processes.

Moreover, it has been found from experimental data \cite{dnbs,dgar} that the 
amplitudes of oscillations of $H_q$ moments increase for more composite 
colliding particles. The anomalous fractal dimensions also differ \cite{ddki}
becoming smaller for $AA$ compared with $p\bar p$ and even more with $e^+e^-$.
Thus, there is no direct similarity of $e^+e^-$ and $p\bar p$-collisions. 

\section{Conclusions}

To conclude, a model of independent pair parton interactions has been proposed. 
It is assumed that hadronic interactions proceed through independent 
parton-parton collisions and each of the binary collisions gives rise to 
a negative binomial distribution of secondary particles with the same
parameters $m$ and $k$. The resulting distribution is described by a weighted 
sum of NBDs whose parameters are equal to the single collision 
values $m$ and $k$ multiplied by the number of pairs.  Thus no new adjustable
parameters appear. Multiple binary parton collisions are assumed to become
more important as energy increases.
A comparison with experimental data at 300, 546, 1000, and 1800 GeV has shown
good agreement. Predictions for the CERN Large Hadron Collider and higher 
energies are presented.
It is demonstrated that asymptotic QCD predictions for $e^+e^-$ multiplicity 
distributions differ from the asymptotic results of the IPPI model for $p\bar p$
processes. Further work on Monte Carlo implementation of this model is in 
progress.

\begin{acknowledgments}

We are grateful to S. Hegyi for correspondence.
This work has been supported in part by the RFBR grants 02-02-16779,
04-02-16445-a, and NSH-1936.2003.2. V.N. is also grateful  
to the Russian Science Support Foundation.
\end{acknowledgments}

\appendix

\section{}

The functions $f_q(k)$ and $\kappa_q(k)$ are $q$th order polynomials of $k$ 
with coefficients determined by the moments $M_r$ with $r\leq q$. Their 
expressions for $q\leq 5$ are as follows:
\begin{widetext}
\begin{eqnarray}
\ff_1 = \kk_1 = M_1 k, \qquad
\ff_2 = M_2 k^2 + M_1 k, \qquad     \kk_2 = (M_2 - M_1^2 ) k^2  + M_1 k,
\nonumber\\
\ff_3 = M_3 k^3 + 3 M_2 k^2 + 2 M_1 k,
\quad
\kk_3 = ( M_3 - 3 M_1 M_2 + 2 M_1^3 ) k^3 + 3 ( M_2 - M_1^2) k^2  + 2  M_1 k,
\nonumber\\
\ff_4 = M_4 k^4 + 6 M_3 k^3 + 11 M_2 k^2 + 6 M_1 k,
\nonumber\\  
\kk_4 = (M_4 - 4 M_1 M_3 + 12 M_1^2 M_2 - 3 M_2^2 - 6 M_1^4 ) k^4
 + 6 (M_3   - 3 M_1 M_2 + 2 M_1^3) k^3 + 11 ( M_2 -  M_1^2 ) k^2  + 6 M_1 k,
\nonumber\\
\ff_5 = M_5 k^5 + 10 M_4 k^4 + 35 M_3 k^3 + 50 M_2 k^2+ 24 M_1 k,
\nonumber\\
  \kk_5 = ( M_5 - 5 M_1 M_4 + 20 M_1^2 M_3 - 60 M_1^3 M_2 + 30 M_1 M_2^2  
            - 10 M_2 M_3 + 24 M_1^5 ) k^5  
         + 10 ( M_4 - 3 M_2^2 - 4 M_1 M_3 
         \nonumber\\
         + 12 M_1^2 M_2 - 6 M_1^4 ) k^4
         + 35 ( M_3 - 3 M_1 M_2 + 2 M_1^3 ) k^3
         + 50 ( M_2 - M_1^2 ) k^2  + 24 M_1 k.
\end{eqnarray}
\end{widetext}

The ratio of the coefficients in front of the leading $k^q$ terms in $\kappa_q$ 
and $f_q$ gives $H_q$ for the Poisson distribution. Thus, in general, $H_q$ 
differs from 0 for a multicomponent Poisson distribution.

The case of one active pair corresponds to $M_r\equiv 1$, and the ordinary 
formula of NBD is restored: $f_q=\Gamma (k+q)/\Gamma (k)$.

To demonstrate the accuracy of the $w_j$ values shown in Table~\ref{tab:table1} 
for the IPPI model, we 
present here their more accurate values and moments for four active parton pairs:

\begin{eqnarray}
w_1 = 0.51879, \ 
w_2 = 0.26914, \ 
w_3 = 0.13963, \ 
\nonumber\\
w_4 = 0.07244, \
M_1 = 1.76571, \ 
M_2 = 4.01103, \ 
\nonumber\\
M_3 = 11.0779, \ 
M_4 = 34.6791, \ 
M_5 = 117.238. \ 
\end{eqnarray}

These values of $w_j$ are larger and those of $M_r$ are smaller than 
the asymptotic ones shown in Eq. (\ref{wm}).\\


\begin{thebibliography}{99}
\bibitem{dche}
I.M. Dremin and D.S. Chernavsky, ZhETF {\bf 38}, 229 (1960). 
\bibitem{afst}
D. Amati, S. Fubini, A. Stanghellini and A. Tonin, Nuovo Cim. {\bf 26}, 896 (1962).
\bibitem{acdr}
V.N. Akimov, D.S. Chernavsky, I.M. Dremin and I.I. Royzen, Nucl. Phys. B {\bf 14}, 285 (1969).
\bibitem{ddun}
I.M. Dremin and A.M. Dunaevsky, Phys. Rep. {\bf 18}, 159 (1975).
\bibitem{gvh}
A. Giovannini and L. Van Hove, Z. Phys. C {\bf 30}, 391 (1986).
\bibitem{giov}
A. Giovannini, Nuovo Cim. A {\bf 10}, 713 (1972). 
\bibitem{lpol}
P.V. Landshoff and J.C. Polkinghorne, Phys. Rev. D {\bf 18}, 3344 (1978).
\bibitem{hump}
B. Humpert, Phys. Lett. B {\bf 131} 461 (1983).
\bibitem{e735}
E735 Collaboration, T. Alexopoulos et al., Phys. Lett. B {\bf 435}, 453 (1998). 
\bibitem{walk}
W.D. Walker, Phys. Rev. D (2004) (to be published).
\bibitem{cstt}
A. Capella, U. Sukhatme, C.I. Tan and J. Tran Thanh Van, Phys. Lett. B {\bf 81}, 
68 (1979).
\bibitem{kter}
A.B. Kaidalov and K.A. Ter-Martirosyan, Phys. Lett. B {\bf 117}, 247 (1982); 
Sov. J. Nucl. Phys. {\bf 40}, 135 (1984).
\bibitem{mwal}
S.G. Matinyan and W.D. Walker, Phys. Rev. D {\bf 59}, 034022 (1999).
\bibitem{ghal}
T.K. Gaisser and F. Halzen, Phys. Rev. Lett. {\bf 54}, 1754 (1985).
\bibitem{gugo}
A. Giovannini and R. Ugoccioni, Phys. Rev. D {\bf 59}, 094020 (1999); 
{\bf 68}, 034009 (2003).
\bibitem{ddki}
E.A. De Wolf, I.M. Dremin and W. Kittel, Phys. Rep. {\bf 270}, 1 (1996).
\bibitem{13}
I.M. Dremin, Phys. Lett. B {\bf 313}, 209 (1993).   
\bibitem{dgar}
I.M. Dremin and J.W. Gary, Phys. Rep. {\bf 349}, 301 (2001).
\bibitem{41}
I.M. Dremin and V.A. Nechitailo, Mod. Phys. Lett. A  {\bf 9}, 1471 (1994);
JETP Lett. {\bf 58}, 881 (1993).
\bibitem{dabg}
I.M. Dremin, V. Arena, G. Boca et al., Phys. Lett. B {\bf 336}, 119 (1994).
\bibitem{sld}
SLD Collaboration, K. Abe et al., Phys. Lett. B {\bf 371}, 149 (1996).
\bibitem{l3}
L3 Collaboration, P. Achard et al., Phys. Lett. B {\bf 577}, 109 (2003).
\bibitem{e73}
E735 Collaboration, F. Turkot et al., Nucl. Phys. A {\bf 525}, 165 (1991).
\bibitem{arn}
UA5 Collaboration, G. Arnison et al., Phys. Lett. B {\bf 118}, 167 (1982).
\bibitem{yuri}
Yu.L. Dokshitzer, Phys. Lett. B {\bf 305}, 295 (1993).
\bibitem{heg}
S. Hegyi, Phys. Lett. B {\bf 387}, 642 (1996); B {\bf 417}, 186 (1998).
\bibitem{pyth}
M. Bertini, L. L\"onnblad and T. Sj\"ostrand, Comput. Phys. Commun. {\bf 134},
365 (2001); see also http://www.thep.lu.se/$\tilde{\;\;}$torbjorn/Pythia.html.
\bibitem{herw}
HERWIG 6.5, G. Corcella, I.G. Knowles, G. Marchesini, S. Moretti, 
K. Odagiri, P. Richardson, M.H. Seymour and B.R. Webber, JHEP {\bf 0101},
010 (2001); see also http://hepwww.rl.ac.uk/theory/seymour/herwig/. 
\bibitem{phob}
PHOBOS Collaboration, B.B. Back et al., nucl-ex/0301017.
\bibitem{dnbs}
I.M. Dremin, V.A. Nechitailo, M. Biyajima and N. Suzuki, Phys. Lett. B
{\bf 403}, 149 (1997).
\end{thebibliography}
\end{document}